\documentstyle[12pt]{ioplppt}

\newcommand{\ol}[1]{\overline{#1}}
\newcommand{\ul}[1]{\underline{#1}}
\newcommand{\what}[1]{\widehat{#1}}
\newcommand{\wtld}[1]{\widetilde{#1}}
\newcommand{\mvec}[1]{\mbox{\boldmath{$#1$}}}
\newcommand{\vx}{\mbox{\boldmath{$x$}}}
\newcommand{\vv}{\mbox{\boldmath{$v$}}}
\newcommand{\vk}{\mbox{\boldmath{$k$}}}
\newcommand{\vu}{\mbox{\boldmath{$u$}}}
\newcommand{\vs}{\mbox{\boldmath{$s$}}}
\newcommand{\psitwo}{\psi_{\mbox{\tiny{II}}}}
\newcommand{\psithree}{\psi_{\mbox{\tiny{III}}}}
\newcommand{\ftwo}{f_{\mbox{\tiny{II}}}}

\begin{document}

\title{Rate increase in chemical reaction and its variance under turbulent equilibrium}

\author{Shunichi Tsug\'e\footnote{e-mail address: shunt@tara.tsukuba.ac.jp}}
\address{Tsukuba Advanced Research Alliance, University of Tsukuba, Tsukuba, 305-8577 Japan\footnote{present address: 790-3 Tohigashi, Tsukuba, 300-2633 Japan}}

\begin{abstract}
As contrast to the Maxwellian distribution, equilibrium distribution of the second kind
or turbulent equilibrium is shown to exist
under tertiary molecular chaos hypothesis to replace the classical binary chaos by
Boltzmann. It is expressed as bimodal Maxwellians each mode differing by plus/minus
root-mean-squares of fluctuations in macroscopic variables.
Chemical reaction rates calculated using the turbulent-equilibrium
are obtained in a closed form, accounting for discrepancy between
experiments and classical theory based on Arrhenius' law that underestimates the burning
rate considerably. The key issue is the correct estimation of the high-energy tail of
the distribution function that is minor in population, yet playing a major role for
reactions with high activation energy.
It is extremely sensitive to turbulence level in the temperature, causing
slightly subcritical molecules to clear the potential barrier
to cause reactions just like quantum particles undergoing
tunnelling effect owing to the uncertainty principle. Variance of the fluctuating
turbulent chemical reaction rate is also calculated, verifying that relative variance
based on the turbulent equilibrium is low, whereas its classical counterpart (Arrhenius)
is pathologically high. A closed set of equations governing reactive turbulent
gases is presented on this sound basis.
\end{abstract}


\section{Introduction --- A perspective on the classical and turbulence- corrected Boltzmann formalism}
It is well-known that each of the transport processes in fluids, namely, diffusion,
heat transfer, momentum transfer due to shearing motion, and chemical reaction is
enhanced by the presense of turbulence. The physical origin that causes such appreciable
hike in those rates is the fractal nature of turbulence\cite{b01}: it is ascribed to drastic
increase in net contact area of adjacent bulk of fluid, taking place through the bounding
surface penetrating into each other area, thereby making it easier to transport mass,
momentum and energy.
It results in drag rise of a flat plate several times\cite{b02} the laminar counterpart,
also in flame propagation velocity elevated by a few ten times\cite{b03} under internal
combustion engine environment(Fig.1). Obviously the latter is a favorable facet of
turbulence characteristies, whereas the former is detrimental from the viewpoint of
transport vehicle technologies. In fact, the evidence that the rate increase due to
turbulence is much higher for mechanism generating thrust rather than drag tempts us to
imagine a midget planet where airplanes would have to install disproportionally huge
engines, under the environment of {\it no} turbulence.
Thus chemical reaction rates, among other transport processes, must be influenced by
turbulence most sensitively, so is worth looking into most detail.

In general there are two aspects of approach to this problem: The one is to pursue
the instantaneous value of a physical quantity $\ul{W}$ which is stochastic
and fractal, using direct numerical simulation (DNS) for example. In principle, any of
these quantities are described in terms of microscopic density advocated by
Klimontovich\cite{b04}, which we will call hereafter the K-formalism. The DNS currently
being used in numerical simulation of fluid flows is the macroscopic version of
the K-formalism\cite{b05}.
It has been applied to problems of combustion science, enabling to compute wiggling
premixed flame front with sufficient resolvability\cite{b06}. The state of the art, however,
is yet to go far to be able to provide information such as turbulent flame speed for
engine designers.

Another approach to turbulent combustion is based on an axiom of statistical mechanics
that claims equivalence of solving for instantaneous value $\ul{W}$ of a
stochastic quantity and for a set of averaged quantities
\begin{equation}\label{eq01}
\{ W,\, \ol{W' \what{W}'},\, \ol{W'\what{W}'\wtld{W}'},\,\cdots \}
\end{equation}
In the above $W=\ol{\ul{W}}$ is the average value,
\begin{equation}\label{eq02}
W' = \ul{W} - W
\end{equation}
is the instantaneous fluctuation subject to $\ol{W'}=0$, and fluctuation correlations
$\ol{W'\what{W}'}$ etc., refer to those between different points in the independent
variable $z,\, \rm{e.g.},\, W=W(z),\, \what{W}=W(\what{z})$, etc. The second
formalism is the
basis on which to found statistical mechanics for example; in fact, if one identifies
$\ul{W}$ with the microscopic density, then its average $W$ represents Boltzmann's
function. So the latter may well be called B-formalism as contrast to the K-formalism as
defined above.

A flow of turbulent gas is {\it dually} stochastic; the one is microscopic (molecular)
level with the average taken with respect to the molecular velocity, whereas the other
is macroscopic (fluid-dynamic) level where the average is taken over a volumelet whose
size is greater than the Kolmogorov scale yet smaller than fluid-dynamic characteristic
length.

If one identifies $W$ with chemical reaction rate, there are three levels of
descriptions: Following the K-formalism $\ul{W}$ is expressed as a sum of on-off
step functions
depending on whether each molecular collision clears the potential barrier to
effect inelastic collisions\cite{b05}.
Next level is the classical B-formalism where the molecular average alone is employed.
Then $W$ is shown to be expressed as Arrhenius' law\cite{b07,b08,b09}. It is coupled with
fractal analysis to estimate turbulent flame surface area to evaluate turbulent burning
rate\cite{b10}. The third stage, namely, the B-formalism to be discussed here is intended
to encompass both molecular and turbulent averaging procedures in its framework.

A series of earlier works along this line\cite{b11,b12,b13} has shed some light on
the conflicts between
the classical Arrhenius kinetics and experimental data attributable to temperature
turbulence using the exact solution that represents simplified
reality.
In Fig.2 is shown shock-tube data of low-temperature (spotty) ignition of hydrogen
oxygen premixture as compared with the classical theory ($\delta T = 0$), also with
turbulence-corrected one, a primitive form\cite{b11} of the present theory. Also is shown
in Fig.3 flame velocity of turbulent premixed gas\cite{b14} as compared with existing
experiment and with renormalization group theory.

Although the earlier theory has thus enabled to predict some features of turbulent
combustion as to their dependence on turbulent intensity without relying on any
empirical parameters, no information has been provided regarding their dependence on the
scale of turbulence. Revised version in the basic formalism has been proposed in
ref. \cite{b15},
with applications to nonreactive turbulence, where full informations on the scale of
turbulence have been made available.

The objective of this paper is three-fold: The one is to show the exact solution for
turbulent equilibrium still valid even without the condition of only temperature
turbulence prevalent. The other is to remodel the previous theory so as to include
scale of turbulence into formalism for reactive flows now at issue.
The third is to show that the B-formalism, an average-oriented formalism, provides
a sound description of turbulent combustion if based on turbulent equilibrium,
which otherwise has been considered as out-of-date.

\section{Turbulent equilibrium}
In paralled with the classical kinetic theory providing molecular basis of nonturbulent
fluid dynamics, molecular description of gases based on the microscopic density
\begin{equation}\label{eq03}
\ul{f}(z)=\sum_{s=1}^N \delta (z-z_s(t))
\end{equation}
has given the ground for the direct numerical simulation of turbulent gases\cite{b05,b16}. In
the above expression, $z=(\vx,\,\vv)$ is a point in the phase space, namely, physical
($\vx$) plus molecular velocity ($\vv$) space, and $\delta$ denotes the (6D) delta
function. This function gives, by definition, the instantaneous number density in the
phase space, so is identified with unaveraged Boltzmann function. It was first applied
to the plasma kinetic theory\cite{b04} where the equation governing $\ul{f}$ is Vlasov's
equation. For monatomic (neutral) ideal gases, the governing equation, derived as an
equation of continuity in the phase space with precise spatial and temporal
resolvability of Hamiltonian mechanics level, proves to be
an unaveraged Boltzmann equation\cite{b16}
\begin{equation}\label{eq04}
B(\ul{f}) \equiv \left( \frac{\partial}{\partial t}+\vv\!\cdot\!\frac{\partial}{\partial \vx}\right) \ul{f} - J(z;\wtld{z})[\ul{f}\,\ul{\wtld{f}}] = 0
\end{equation}
where $J$ is the classical collision integral operator
\begin{equation}\label{eq05}
J(z; \wtld{z})[\ul{f}\,\ul{\wtld{f}}] = \left[ \int (f'\wtld{f}'-f\wtld{f})V\,d\Omega\,d\wtld{\vv} \right]_{\wtld{\vx}=\vx}
\end{equation}
In the above $\ul{\wtld{f}} = \ul{f}(\wtld{z})$ is the microscopic density of the
collision
partner molecule $\wtld{z}$, $\wtld{f}'=f(\wtld{z}')$ likewise with $\what{z}'$
specifying the pre-collision state of the collision partner leading to $\wtld{z}$ after
the collision, $V=|\wtld{\vv}-\vv|$, and $d\Omega$ is the differential collision-cross
section of the molecular encounter.

The microscopic density, being averaged over a space with a number of unidentifiable
molecules included, defines the Boltzmann function $f$;
\begin{equation}\label{eq06}
f = \ul{\ol{f}}\,.
\end{equation}
This is where the statistical concept is introduced for a quantity of deterministic
mechanics.
Similarly, two-point correlation function $\psi(z, \wtld{z})$ is defined using the same
averaging concept as
\begin{equation}\label{eq07}
\psi(z, \wtld{z}) = \ol{\ul{f}\,\ul{\wtld{f}}}-f\wtld{f}
\end{equation}
Higher order correlations can be defined likewise.

The classical kinetic theory is founded on Eq.(4) averaged over the same space as
(6) together with the Boltzmann's (binary) molecular chaos hypothesis
\begin{equation}\label{eq08}
\psi = 0
\end{equation}
that is the Boltzmann equation in the classical sense;
\begin{equation}\label{eq09}
\left( \frac{\partial}{\partial t}+\vv\!\cdot\!\frac{\partial}{\partial \vx}\right) f - J(z;\wtld{z})[f\wtld{f}] = 0
\end{equation}
Under the equilibrium condition ($\partial /\partial t=0,\,\partial /\partial \vx = 0$)
the equation reduces to
\begin{equation}\label{eq10}
J(z;\wtld{z})[f\wtld{f}]=0
\end{equation}
whose solution is known as the Maxwellian distribution
\begin{equation}\label{eq11}
\left. \begin{array}{l}
\displaystyle f=f_0=\frac{n}{(2\pi c^2)^{3/2}} \exp \left(-\frac{\mvec{w}^2}{2c^2}\right) \vspace*{2mm} \\
\mvec{w} \equiv \vv-\vu \vspace*{2mm} \\
c=({\cal R}T/{\cal M})^{1/2}
\end{array} \right\}
\end{equation}
where $n,\, \vu,\, T$ are the number density, fluid velocity and temperature, ${\cal R}$
and ${\cal M}$ denote the universal gas constant and molar weight, respectively.

If a gas is turbulent, fluid variables exhibit (long-range) correlations, so the
classical binary chaos (8) ceases to hold. Then Eq.(9) is to be replaced with
\begin{equation}\label{eq12}
\ol{B(\ul{f})}=\left( \frac{\partial}{\partial t}+\vv\!\cdot\!\frac{\partial}{\partial \vx}\right) f-J(z;\wtld{z})[f\wtld{f}+\psi(z,\wtld{z})]=0
\end{equation}
Thus the equation is not closed at this level: The equation of next step to govern
$\psi$ is\cite{b15}
\begin{equation}\label{eq13}
\ol{f' B(\ul{\what{f}})+B(\ul{f})\what{f}'}=0~~~~(f' \equiv \ul{f}-f)
\end{equation}
which is shown to be identical with the two-particle hierarchy equation of the
BBGKY theory\cite{b17}. The actual form of (13) reads
\begin{equation}\label{eq14}
\begin{array}{l} \displaystyle
\left(\frac{\partial}{\partial t}+\vv\!\cdot\!\frac{\partial}{\partial \vx}+\what{\vv}\!\cdot\!\frac{\partial}{\partial \what{\vx}}\right)\psi (z,\what{z})=J(z;\wtld{z})[f \psi(\what{z}, \wtld{z})+\wtld{f} \psi (z,\what{z})\vspace*{2mm}\\
\hspace*{15mm}+\psithree(z,\what{z},\wtld{z})]+J(\what{z};\wtld{z})[\what{f}\psi (z,\wtld{z})+\wtld{f}\psi(z,\wtld{z})+\psithree(z,\what{z},\wtld{z})]
\end{array}
\end{equation}

The simplest possible truncation of the chain of equations at this level is to put
\begin{equation}\label{eq15}
\psithree =0
\end{equation}
namely, to invoke tertiary molecular chaos to replace the classical binary chaos (8).

Eq.(14) subject to condition (15) turns out to be separable into those for respective
variables in terms of the wave number $\vk$,
\begin{equation}\label{eq16}
\psi(z,\what{z})={\rm R.P.} l^3 \int \phi(z,\vk) \phi^*(\what{z},\vk)d\vk
\end{equation}
where symbol $*$ denotes complex conjugate, $l$ denotes a length characteristic of
the flow geometry and R.P. stands for the real part.
Upon its substitution into (14) we have the equation governing $\phi$ as
\begin{equation}\label{eq17}
\left. \begin{array}{l}
i \omega(\vk) \phi = \Omega_0(\phi) \vspace*{2mm}\\
\displaystyle \Omega_0(\phi) \equiv \left(\frac{\partial}{\partial t}+\vv\!\cdot\!\frac{\partial}{\partial \vx}\right) \phi-J(z;\wtld{z})[f\wtld{\phi}+\phi\wtld{f}]
\end{array} \right\}
\end{equation}
where $\omega(\vk)$ is the separation constant having dimension of the frequency, and
connected with the wave number through phase velocity $\mvec{V}_p$ as
\begin{equation}\label{eq18}
\omega=\mvec{V}_p\!\cdot\!\vk
\end{equation}
>From Eqs.(17) and (18) the following relationship is seen to hold;
\begin{equation}\label{eq19}
\phi^*(\vk)=\phi(-\vk)
\end{equation}

A remark should be mentioned on a renovation that has brought the scale of turbulence
into formalism: In the older theory variable separation (16) is effected in the
domain of frequency $\omega$ in place of wave number $\vk$. Information on the
scale of turbulence is missed in this form, which may be retrieved in the
following way\cite{b15}:
Here turbulence is characterized as a wave with frequency $\omega$ and wave number
$\vk$ obeying dispersion relationship (18). We may assume a plane wave
\begin{equation}\label{eq20}
\phi = e^{i\vk\cdot\vx}\Phi(z,\,\vk)
\end{equation}
and discuss its amplitude $\Phi$.
If further we introduce Fourier transform
\begin{equation}\label{eq21}
\Phi(z,\,\vk)=\frac{1}{(2\pi)^3}\int_{-\infty}^\infty e^{-i\vk\cdot\vs}F(z,\,\vs)\,d\vs
\end{equation}
to work with `eddy' space $\vs$, and employ separation rule written
in terms of wave number $\vk$, namely after (16), a simple formula follows:
\begin{equation}\label{eq22}
\psitwo(z,\,\what{z})=\frac{1}{(2\pi l)^3}\int_{-\infty}^\infty F(z,\,\vs)F(\what{z},\,\vs+\mvec{r})\,d\vs
\end{equation}
where $\mvec{r}=\what{\vx}-\vx$ is the distance between the two space points.
Any informations on scale of turbulence can be obtained from this correlation
formula by taking fluid moment of (22), for example, the velocity-velocity
fluctuation correlation as
\begin{equation}\label{eq23}
\ol{u_j'(\vx)\mvec{u}_l'(\vx+\mvec{r})}=(\rho\what{\rho})^{-1}\int w_j\what{w}_l\psitwo \,d\vv\,d\what{\vv}
\end{equation}
where $\mvec{w}$ has been defined by (11).

It is a well-posed problem to seek the equilibrium solution of the new set of equations
(12) and (17).
It stands for a `turbulent equilibrium' solution to substitute the local Maxwellian for
nonturbulent gases. This is to solve simultaneous integral equations
\begin{equation}\label{eq24}
J(z;\wtld{z})[f\wtld{f}+\int\phi\,\wtld{\phi}^*\, d\vk]=0
\end{equation}
\begin{equation}\label{eq25}
J(z;\wtld{z})[f\wtld{\phi}+\phi\wtld{f}]=0
\end{equation}
The exact solution under the assumption that turbulence
prevails only in the temperature
has been obtained in a closed form by Sagara\cite{b12} by summing an
infinite series in the temperature fluctuation $\delta T=(\ol{T^{'2}})^{1/2}$.
The same result has been reached using much simpler method\cite{b13}.

We can show that the exact solution also exists for realistic turbulence where the
fluctuations in number density $\delta n=(\ol{n^{'2}})^{1/2}$ and fluid velocity
$\delta u_j=(\ol{u_j^{'2}})^{1/2}$ (no summation convention here) are also prevalent. Put
\begin{equation}\label{eq26}
f=f^0=\frac12 (f_0^++f_0^-)
\end{equation}
\begin{equation}\label{eq27}
\phi=\phi^0=\frac12 K(\vk)(f_0^+-f_0^-)
\end{equation}
where $f_0$ is the local Maxwellian defined by (11), and $f_0^\pm$ is given as follows;
\begin{equation}\label{eq28}
\left. \begin{array}{ccl}
f_0^\pm &=& \displaystyle  \frac{n^\pm}{(2\pi c^{\pm 2})^{3/2}} \exp \left(-\frac{w_k^{\pm2}}{2c^{\pm 2}} \right) \vspace*{2mm}\\
n^\pm &=& n \pm \delta n \vspace*{2mm}\\
w_j^\pm &=& v_j-u_j^\pm \vspace*{2mm}\\
u_j^\pm &=& u_j \pm \delta u_j \vspace*{2mm}\\
c^{\pm 2} &=& R(T \pm \delta T)\vspace*{2mm}\\
R &=& {\cal R}/{\cal M}
\end{array} \right\}
\end{equation}
It is easily confirmed that the form of (26) is designed to be consistent with
definitions of the average quantity $(n,\, \vu,\, T)$, and the form of (27) with that of
the variance $(\delta n,\, \delta\vu,\, \delta T)$ of each quantity provided that the
following relationship holds;
\begin{equation}\label{eq29}
\int_{-\infty}^{\infty} KK^* d\vk =1
\end{equation}
Direct substitution of (26) and (27) into (24) and (25) reads, respectively,
\begin{equation}\label{eq30}
\left. \begin{array}{c}
\displaystyle \frac12 J(z;\what{z})[f_0^+\what{f}_0^++f_0^-\what{f}_0^-]=\frac12 (J^++J^-) \vspace*{2mm} \\
\displaystyle \frac{K}{2} J(z;\what{z})[f_0^+\what{f}_0^+-f_0^-\what{f}_0^-]=\frac{K}{2} (J^+-J^-)
\end{array}\right\}
\end{equation}
where $J^\pm$ has been defined as
\begin{equation}\label{eq31}
J^\pm = J(z;\what{z}) [f_0^\pm \what{f}_0^\pm ]
\end{equation}
This integral vanishes in view of the fact that $J(z;\what{z})[f_0\what{f}_0]=0$, also that the bilinear Maxwellians in (31) consist of the same family in
fluid parameters (Q.E.D.).

The deduction above claims that the state of turbulent equilibrium is represented by
the Boltzmann function of the form (26), along with correlation function
\begin{equation}\label{eq32}
\psi(z,\what{z}) = \frac14 (f_0^+ - f_0^-)(\what{f}_0^+-\what{f}_0^-)
\end{equation}
or the two-point distribution function
\begin{equation}\label{eq33}
\begin{array}{rcl}
\ftwo (z,\what{z}) &\equiv& f\what{f}+\psi(z,\what{z}) \vspace*{2mm} \\
 &=& \displaystyle \frac12 (f_0^+\what{f}_0^++f_0^-\what{f}_0^-)
\end{array}
\end{equation}

At a glance difference between the two Boltzmann functions $f_0$ of (3) and $f^0$ of
(26) looks trivial, because the difference is of the first order smallness in $\delta$.
The point at issue, however, is that they differ appreciably in the population of high
energy molecules that play a crucial role in chemical reactions, particularly when the
activation energy is high. In Fig.4 is shown the population density of molecules
having absolute molecular velocity $v$ with turbulence only in the temperature.
The Maxwellian distribution $(\delta T=0)$ is seen to have lean high energy tail,
namely, underestimates population of high energy molecules compared to reality of
turbulence.

\section{Turbulent chemical reactions}
Chemical reaction rate of elementary reaction of the form
$$
[{\rm a}]+[{\rm b}]\rightarrow [{\rm c}]+[{\rm d}]~~~~([~]\,;\,{\rm chemical~symbol})
$$
usually obeys Arrhenius' law;
\begin{equation}\label{eq34}
W=An_{\rm a}n_{\rm b}\exp(-E/{\cal R}T)
\end{equation}
where $n_{\rm a}$ and $n_{\rm b}$ denote partial number densities of reactants $[{\rm a}]$ and $[{\rm b}]$,
$A$ is the frequency factor and $E$ is the activation energy per mol.
Rate law (34) is warranted on the classical kinetic theory using a simple
collision model. A most clearcut model among various ones proposed in
late 40's through 50's is due to Present\cite{b09} : Let the intermolecular
potential of a molecule [a] relative to a molecule [b] be repulsive, of a
volcano shape having a crater with radius $r_0$ at altitude $\epsilon$.
Let also the locus of the [b] molecule relative to the [a] molecule written
in axisymmetric polar
coordinate be $[r(t), \chi(t)]$ where $\chi$ is the angle between the symmetry
axis parallel to $\mvec{V}_{\rm ba}=\vv_{\rm b}-\vv_{\rm a}$ and the position vector
$\mvec{r}=\vx_{\rm b}-\vx_{\rm a}$ then we have
\begin{equation}\label{eq35}
\left. \begin{array}{l}
r^2 \dot{\chi} = b V \vspace*{2mm}\\
(m^\dagger /2)(\dot{r}^2+r^2\dot{\chi}^2)+\epsilon(r)=(m^\dagger /2)V^2
\end{array} \right\}
\end{equation}
relationships representing conservation of angular momentum and of
energy, where $\dot{\chi}=dx/dt$ etc., $m^\dagger=(m_{\rm a}^{-1}+m_{\rm b}^{-1})^{-1}$
is the reduced mass, $b$ is the impact parameter, and $\epsilon (r)$ denotes the
intermolecular potential. Since the radius of the
closest approach $r=r_c$ is given by the condition $dr/d\chi =0$, the critical
collision that separates elastic from inelastic collisions is such that the
closest approach occurs on the ridge line of the crater ; $r_c=r_0$. This
condition determines the critical impact parameter $b_c(V)$ from (35) as
\begin{equation}\label{eq36}
\frac{b_c(V)}{r_0}=\left( 1-\frac{2\epsilon}{m^\dagger V^2} \right)^{1/2}
\end{equation}

Instantaneous reaction rates for respective species are expressed in the
following form,
\begin{equation}\label{eq37}
\begin{array}{l}
\ul{W}_{\rm a} = -m_{\rm a} \ul{W},~\ul{W}_{\rm b}=-m_{\rm b}\ul{W},~\ul{W}_{\rm c}=m_{\rm c}\ul{W},~\ul{W}_{\rm d}=m_{\rm d}\ul{W} \vspace*{2mm}\\
\displaystyle \ul{W}=\int_{\rm inel} \ul{f}(z_{\rm a})\ul{f}(z_{\rm b})V\,d\Omega_{\rm ab}\,d\vv_{\rm a}\,d\vv_{\rm b}
\end{array}
\end{equation}
Utilizing the following formula
$$
\int_{\rm inel} d\Omega_{\rm ab}=\pi b_c^2=\pi r_0^2\left( 1-\frac{2\epsilon}{m^\dagger V^2}\right)
$$
and transforming integral variables from $(\vv_{\rm a},\,\vv_{\rm b})$ to $(\mvec{U},\,\mvec{V})$ defined by
\begin{equation}\label{eq38}
\left. \begin{array}{rcl}
\mvec{U} &=& (m_{\rm a}\vv_{\rm a} +m_{\rm b}\vv_{\rm b})/(m_{\rm a}+m_{\rm b}) \vspace*{2mm}\\
\mvec{V} &=& \vv_{\rm b}-\vv_{\rm a}
\end{array}\right\}
\end{equation}
we have for (33)
\begin{equation}\label{eq39}
\ul{W}=\pi r_0^2 \int_{V_c}^\infty d\mvec{V}\,V\left(1-\frac{V_c^2}{V^2}\right)\int_{-\infty}^{\infty}\ul{f}(z_{\rm a})\ul{f}(z_{\rm b})d\mvec{U}
\end{equation}
where
\begin{equation}\label{eq40}
V_c=(2\epsilon / m^\dagger)^{1/2}
\end{equation}
denotes the critical relative velocity for a head-on collision.

Chemical reaction rate in classical sense is obtained from (39), upon
averaging and employing classical chaos (8), as
\begin{equation}\label{eq41}
W=\pi r_0^2 \int_{V_c}^\infty d\mvec{V}\,V\left(1-\frac{V_c^2}{V^2}\right)\int_{-\infty}^{\infty}f(z_{\rm a})f(z_{\rm b})d\mvec{U}
\end{equation}
If the Maxwellian equilibrium (11) is employed Eq.(41) is integrated out to
give
\begin{equation}\label{eq42}
\left. \begin{array}{l}
W_0(T,\,n_{\rm a},\,n_{\rm b})=A n_{\rm a} n_{\rm b} e^{-\epsilon / kT} \vspace*{2mm}\\
A=(8 kT/\pi m^\dagger)^{1/2} \pi r_0^2
\end{array} \right\}
\end{equation}
that is nothing but Arrhenius law (34) with activation energy $E$ and the
frequency factor $A$ written in microscopic parameters.

It should be noted that turbulence counterpart of the above deduction faces
with difficulty if conducted phenomenologically, whereas it is simply
straightforward if one follows the molecular approach as employed here.
In fact, if one assumes that the instantaneous reaction rate is Arrhenius
form (34)
$$
\ul{W}=\ul{A}\,\ul{n}_{\rm a}\ul{n}_{\rm b} \exp(-E/{\cal R}\ul{T})
$$
and its average given by
\begin{equation}\label{eq43}
W=\ol{\ul{A}\,\ul{n}_{\rm a}\ul{n}_{\rm b} \exp(-E/{\cal R}\ul{T})}
\end{equation}
Decomposing the r.h.s. term under average into means and fluctuations we have
$$
W=An_{\rm a}n_{\rm b} \ol{(1+O(n_{\rm a}',\,n_{\rm b}'))e^{-\beta+\beta^2\tau-\beta^3\tau^2+\cdots}}
$$
where $\beta=E/{\cal R}T$ is the Zel'dovich number and $\tau={\cal R}T'/E$ stands for
the temperature fluctuation. Leading term in the Taylor expansion of the
expression under average sign gives for $\beta \gg 1$,
$$
\ol{\frac{d^nW}{d\tau^n}\frac{\tau^n}{n!}} \sim \frac{\beta^n}{n!} \ol{(T')^n}
$$
which implies that convergence of the series is extremely slow since the
Zel'dovich number for combustion reaction is of $O(10^1)$. For example we
need to take 23 terms before $\beta^n/n!$ diminishes to unity for
$\beta=10$. It also means that we need to know 22 consecutive
self correlations in the temperature fluctuation. Needless to say, therefore,
that the first term (Arrhenius) or a few-term approximation is far from
satisfactory.

On the other hand, if one adopts microscopic approach, namely, starts with
the averaged version of (39) insted of (43);
\begin{equation}\label{eq44}
W=\pi r_0^2\int_{V_c}^\infty d\mvec{V} V\left(1-\frac{V_c^2}{V^2}\right)
\int_{-\infty}^\infty \ol{\ul{f}_{\rm a}\ul{f}_{\rm b}} d\mvec{U}
\end{equation}
and employs turbulent equilibrium (33) for $\ol{\ul{f}_{\rm a}\ul{f}_{\rm b}} \!
\equiv \!\ftwo$,
a simple calculation leads immediately to the final expression
\begin{equation}\label{eq45}
W^0=\frac12 (W_0^++W_0^-)
\end{equation}
with
\begin{equation}\label{eq46}
W_0^\pm \equiv W_0(T\pm \delta T,\,n_{\rm a}\pm\delta n_{\rm a},\,n_{\rm b}\pm \delta n_{\rm b})
\end{equation}
where $W_0$ has been defined by (42). Formula (45) for the turbulent reaction
rate turns out to be a bimodal Arrhenius: It allows us to draw a intuitive
picture that a half of the chemical reaction takes place under temperature
elevated and population of reactant molecules denser by their root-mean-square turbulent
intensity, and another half under those quantities lowered by the same amount.

The bimodal Arrhenius low (45) is the full turbulence version derived on the
basis of exact solution of Eqs. (24) and (25), supplementing the previous
one\cite{b11,b12,b13}, where turbulence prevails only in the temperature.
As already sketched it is obvious that anomalous increase in the reaction
rate is overwhelmingly due to the temperature turbulence that feeds extra
high energy molecules capable of reaction
(see Fig.4). The first term of (45) looks as if the potential barrier
is lowered by the factor of $(1+ \delta T/T)^{-1}$; a macroscopic equivalent
of tunnelling effect of quantum mechanics where the potential barrier is
lowered by a factor of Planck's constant due to the uncertainty principle.

On the other hand, fluctuation in the reactant number density does not have
such effects on the reaction rate beyond the first order smallness.
However, when coupled with the temperature effect, it leaves with a
possible account for below-critical lean combustion that actually occurs
in the operation of the direct injection engines (GDI).

Rate law (45) also tells us that turbulence in the fluid velocity has no
influence on the reaction rate (as it should not). In fact, the collision
integral for a pair of molecules about to collide are not influenced by
the fluctuation in the fluid velocity, because they are aboard on the
`same boat' in the turbulent ocean. Its effects on turbulent flame velocity
are exclusively via other transport processes, namely, turbulent diffusion and
turbulent heat transfer. (See Table 1, next section).

\section{Reaction rate variance}
It is an experimental evidence that instantaneous chemical reaction rate
$\ul{W}=W+W'$ at a fixed point undergoes fluctuation over a wide range, therefore,
quality of a statistical theory employing any average concepts hinges crucially on
reliable estimate for its variance around the average $W$.

This issue again is beyond the reach of phenomenologies and we have to
address to statistical approach as developed in the preceding section.
To be remarked at this
point is that what appear in the governing equations are not the variance
itself, but the fluctuation correlations $\ol{Z'W'}$ where $Z$ stands for
fluid variables such as velocity, temperature, etc..

The equation governing $W'$ is obtained from (39) by separating it into
mean and fluctuating parts. We have, then,
\begin{equation}\label{eq47}
W'=I_{\rm inel}\left[ f'(z_{\rm a})f(z_{\rm b})+f(z_{\rm a})f'(z_{\rm b})+f'(z_{\rm a})f'(z_{\rm b})-\ol{f'(z_{\rm a})f'(z_{\rm b})}\right]
\end{equation}

\noindent
where we have defined
$$
I_{\rm inel}[\Phi (z_{\rm a},\, z_{\rm b})] \equiv \pi r_0^2 \int_{V_c}^\infty d\mvec{V} V\left(1-\frac{V_c^2}{V^2}\right)\int_{-\infty}^\infty \Phi(z_{\rm a},\,z_{\rm b})d\mvec{U}
$$
Since, in general, fluid variable fluctuation is expressed as a moment of
fluctuation of the Boltzmann function in the velocity space as
\begin{equation}\label{eq48}
Z'=\int_{-\infty}^\infty \zeta(z_\alpha)f'(z_\alpha)d\vv_\alpha
\end{equation}
we have, from (43) and (44),
\begin{equation}\label{eq49}
\int_{-\infty}^\infty \zeta d\vv_\alpha \left\{ \ol{W'f_\alpha'}-I_{\rm inel}\left[ \ol{f_{\rm a}' f_\alpha'}f_{\rm b}+f_{\rm a}\ol{f_\alpha' f_{\rm b}'}+\ol{f_{\rm a}'f_\alpha'f_{\rm b}'}\right]\right\} =0
\end{equation}
Let the variable separation be invoked after (16) as
\begin{equation}\label{eq50}
\ol{W'f_\alpha'}={\rm R.P.}l^3 \int_{-\infty}^\infty g_W(\vk)\phi_\alpha^*(\vk)d\vk
\end{equation}
and
$$
\begin{array}{rcl}
\ol{f_{\rm a}'f_\alpha'f_{\rm b}'}&=&\displaystyle {\rm R.P.}l^9 \int \phi_{\rm a}(\vk_{\rm a})\phi_\alpha(\vk_\alpha)\phi_{\rm b}(\vk_{\rm b})\delta(\vk_{\rm a}+\vk_\alpha+\vk_{\rm b})d\vk_{\rm a}\,d\vk_\alpha\,d\vk_{\rm b} \vspace*{2mm}\\
 &=&\displaystyle {\rm R.P.}l^6 \int \phi_\alpha^*(\vk)\phi_{\rm a}(\wtld{\vk})\phi_{\rm b}(\vk-\wtld{\vk})d\vk\,d\wtld{\vk}
\end{array}
$$
where, in the last row, $-\vk_\alpha$ is replaced with $\vk$, also $\vk_{\rm a}$ with $\wtld{\vk}$,
and condition (19) has been made use of. Then Eq.(49) leads to
\begin{equation}\label{eq51}
\begin{array}{l}
\displaystyle \left. \int_{-\infty}^\infty d\vk \phi_\alpha^*(\vk)\right\{g_W(\vk)-I_{\rm inel}[\phi_{\rm a}(\vk)f_{\rm b}+f_{\rm a}\phi_{\rm b}(\vk) \vspace*{2mm}\\
\hspace*{26mm}\displaystyle \left. +{\rm R.P.} l^3 \int_{-\infty}^\infty d\wtld{\vk}\phi_{\rm a}(\vk-\wtld{\vk})\phi_{\rm b}(\wtld{\vk})d\wtld{\vk}]\right\}=0
\end{array}
\end{equation}
If one decomposes $g_W$ and $\phi$ into a plane wave after (20)
\begin{equation}\label{eq52}
\left. \begin{array}{rcl}
g_W &=& e^{i\vk\cdot\vx}G_W \vspace*{2mm}\\
\phi &=& e^{i\vk\cdot\vx}\Phi
\end{array} \right\}
\end{equation}
and deals with its amplitude $G_W$ and $\Phi$, or its Fourier-transform
version
\begin{equation}\label{eq53}
\left. \begin{array}{rcl}
G_W(\vx,\, \vk) &=& \displaystyle \frac{1}{(2\pi l)^3} \int_{-\infty}^\infty d\vs\, e^{-i\vk\cdot\vs}q_W(\vx,\,\vs) \vspace*{2mm}\\
\Phi(\vx,\,\vk) &=& \displaystyle \frac{1}{(2\pi l)^3} \int_{-\infty}^\infty d\vs\, e^{-i\vk\cdot\vs}F(\vx,\,\vs)
\end{array} \right\}
\end{equation}
we have two alternatives for the reaction rate fluctuation from (51),
\begin{equation}\label{eq54}
G_W(\vx,\,\vk)=I_{\rm inel}\left[\Phi_{\rm a}f_{\rm b}+f_{\rm a}\Phi_{\rm b}+{\rm R.P.} l^3 \int_{-\infty}^\infty d\wtld{\vk}\Phi_{\rm a}(\vk-\wtld{\vk})\Phi_{\rm b}(\wtld{\vk})d\wtld{\vk}\right]
\end{equation}
\begin{equation}\label{eq55}
q_W(\vx,\,\vs)=I_{\rm inel}[F_{\rm a}f_{\rm b}+f_{\rm a}F_{\rm b}+F_{\rm a}F_{\rm b}]
\end{equation}
depending on which of the wave number $(\vk)$ or the eddy $(\vs)$ spaces is
more convenient to work with.

The actual form of $q_W$ associated with turbulent equilibrium (26) and (27)
is straightforward. In fact, in view of (52) and (53), we have
\begin{equation}\label{eq56}
F_\alpha^0(z,\,\vs)=S(\vx,\,\vs)\frac{f_{0\alpha}^+ - f_{0\alpha}^-}{2}
\end{equation}
with
\begin{equation}\label{eq57}
S(\vx,\,\vs)=(2\pi l)^{3/2}\frac{q_0}{\displaystyle \left(\int_{-\infty}^\infty q_0^2\,d\vs\right)^{1/2}}
\end{equation}
where $q_0$ is related to variance of the density fluctuation by
\begin{equation}\label{eq58}
(\delta \rho)^2 = \frac{1}{(2\pi l)^3} \int_{-\infty}^\infty q_0^2\,d\vs
\end{equation}
Utilizing (56) in (55) we have
\begin{equation}\label{eq59}
q_W(\vx,\,\vk)=\frac{S}{2}\left[ W_0^+-W_0^-+\frac{S}{2}(W_0^++W_0^--W_0^{+-})\right]
\end{equation}
where
\begin{equation}\label{eq60}
\hspace*{-1mm}\begin{array}{l}
\displaystyle W_0^{+-}=A\left\{(n_{\rm a}+\delta n_{\rm a})(n_{\rm b}-\delta n_{\rm b})\exp\left[-E/{\cal R}\left(T+\frac{m_{\rm a}-m_{\rm b}}{m_{\rm a}+m_{\rm b}} \delta T \right)\right]\right. \vspace*{2mm}\\
\hspace*{17mm} \displaystyle +\left. (n_{\rm a}-\delta n_{\rm a})(n_{\rm b}+\delta n_{\rm b})\exp\left[-E/{\cal R}\left(T-\frac{m_{\rm a}-m_{\rm b}}{m_{\rm a}+m_{\rm b}} \delta T \right)\right]\right\}
\end{array}
\end{equation}

Once we have obtained explicit expression for $q_W$ variance of the reaction rate
can be calculated from (see Eq.(50))
\begin{equation}\label{eq61}
\begin{array}{rcl}
\ol{W^{'2}} &=& \displaystyle {\rm R.P.}l^3 \int_{-\infty}^\infty g_Wg_W^* \,d\vk \vspace*{2mm}\\
 &=& \displaystyle \frac{1}{(2\pi l)^3}\int_{-\infty}^\infty q_W^2 \,d\vs
\end{array}
\end{equation}
where (52), (53) have been taken into account.
In Fig.5a is shown the relative variance $(\ol{W^{'2}})^{1/2}/W_0$ of
reaction rate around classical Arrhenius, which amounts to quantities exceeding
$O(1)$ by far, indicating extraordinary fluctuation intensity to be observed
by a fixed-point measurement. Also is shown in Fig.5b the same ratio for
turbulent reaction rate (41). In this case the relative variance stays
within sound realm, and approaches unity with increase in turbulence in the
temperature and with Zel'dovich number. This is a sign of stochastic signals
characteristic of on-off type.

\section{Phenomenology-coupled Boltzmann formalism on turbulent equilibrium}

Statistical description of stochastic phenomena where relative variance far
exceeds unity is pathological. It is for this reason that a formalism using
the averaged Arrhenius law fails in describing turbulent combustion.
However, as asserted in the preceding section, such formalism using the
bimodal law as the averaged reaction rate deserves reconsideration.
For this purpose it is advisable to follow the recipe prepared for nonreactive
turbulence\cite{b15}: We start with phenomenological equations generalized to
reactive gases written in instantaneous quantities:
\begin{equation}\label{eq62}
\ul{\Lambda}_0 \equiv \frac{\partial \ul{\rho}}{\partial t}+\frac{\partial \ul{m}_r}{\partial x_r} = 0
\end{equation}
\begin{equation}\label{eq63}
\ul{m}_r \equiv \ul{\rho}\,\ul{u}_r
\end{equation}
\begin{equation}\label{eq64}
\ul{\Lambda}_j \equiv \frac{\partial m_j}{\partial t}+\frac{\partial}{\partial x_r}\left( \frac{\ul{m}_j\ul{m}_r}{\ul{\rho}}+\ul{p}\delta_{jr}+\ul{p}_{jr}\right)=0
\end{equation}
\begin{equation}\label{eq65}
\ul{p}_{jr} \equiv -\ul{\mu} \left[ \frac{\partial}{\partial x_j}\left( \frac{\ul{m}_r}{\ul{\rho}} \right)+\frac{\partial}{\partial x_r}\left(\frac{\ul{m}_j}{\ul{\rho}}\right)-\frac23 \delta_{jr}\frac{\partial}{\partial x_k} \left( \frac{\ul{m}_k}{\ul{\rho}}\right) \right]
\end{equation}
\begin{equation}\label{eq66}
\hspace*{-3mm} \left. \begin{array}{l}
\displaystyle \ul{\Lambda}_4 \equiv \frac{\partial \ul{E}}{\partial t}+\frac{\partial}{\partial x_r}\left(\ul{m}_r \ul{H}+\ul{Q}_r \right)+\sum e_\alpha^0 \ul{W}_\alpha =0 \vspace*{2mm}\\
\displaystyle \ul{E} \equiv \ul{\rho}\,\ul{e}+\frac{\ul{m}_k^2}{2\ul{\rho}},~\ul{H} \equiv \ul{E}+\ul{p}
\end{array}\right\}
\end{equation}
\begin{equation}\label{eq67}
\ul{Q}_r \equiv -\ul{\lambda}\, \frac{\partial}{\partial x_r}\left( \frac{\ul{p}}{\ul{\rho}R}\right)
\end{equation}
\begin{equation}\label{eq68}
\hspace*{-3mm}\left. \begin{array}{l}
\displaystyle \ul{\Lambda}_{\rm a} \equiv \frac{\partial \ul{\rho}_{\rm a}}{\partial t}+\frac{\partial}{\partial x_r}\left( \ul{\rho}_{\rm a}\ul{u}_r +\ul{M}_{{\rm a},\,r}\right) -\ul{W}_{\rm a}=0 \vspace*{2mm}\\
\ul{Y}_{\rm a} \equiv \ul{\rho}_{\rm a}/\ul{\rho}
\end{array} \right\}
\end{equation}
\begin{equation}\label{eq69}
\ul{M}_{{\rm a},\,r} \equiv -\ul{\rho}\,\ul{D}_{\rm a}\frac{\partial \ul{Y}_{\rm a}}{\partial x_r}
\end{equation}
In the above, $\ul{\rho},\,\ul{m}_j,\,\ul{p}$ and $\ul{e}$ denote density,
massflux density, pressure and specific internal
energy (thermal part only), $\ul{p}_{jr}$, $\ul{Q}_r$ and $\ul{M}_{{\rm a}r}$
are viscous stress, heat flux density and diffusion mass flux density for
species a, and $\ul{\rho}_{\rm a}$, $\ul{Y}_{\rm a}$ and $e_{\rm a}^0$ are
partial density, mass fraction and zero-point enthalpy of species a,
respectively. Also, $\ul{\mu}$, $\ul{\lambda}$ and $\ul{D}_{\rm a}$ are
coefficients of viscosity, thermal conductivity and diffusion for species a,
respectively. Note that all the equations have the form in accordance with
a rule (rule I) that they are written in terms of quantities proportional
to the density. This is to use mass-flux density instead of fluid velocity,
pressure instead of temperature, and internal energy per unit of volume
instead of specific internal energy.

A few words must be mentioned about modified Fick's law (69) that is not
exact consequence of the kinetic theory of gases as distinct from other
molecular transport relationships. The `effective' diffusion coefficient
$D_{\rm a}$ is expressed in terms of binary diffusion coefficient for
each pair of constituents as \cite{b18}
\begin{equation}\label{eq70}
\begin{array}{rcl}
D_{\rm a} &=& \displaystyle (1-Y_{\rm a})/ \sum_{\rm b} D_{\rm ab}^{-1}Y_{\rm b}\vspace*{2mm}\\
 &=& \displaystyle 1/\sum_{\rm b} D_{\rm ab}^{-1} Y_{\rm b}'
\end{array}
\end{equation}
where $Y_{\rm b}'=Y_{\rm b}/(1-Y_{\rm a})$ is mass fraction of species b of
the mixture with species a excepted. The approximate formula (69) with (70)
warrants validity for practical use except a minor flaw that condition
$\displaystyle \sum_{\rm a} \ul{M}_{\rm a, r} = 0$ is not exactly met.

Equations for the average are provided by
\begin{equation}\label{eq71}
\Lambda_\alpha \equiv \ul{\ol{\Lambda}}_\alpha =0 ~~~~(\alpha\,;\,0,\,j,\,4,\,{\rm a},\,{\rm b},\,\cdots)
\end{equation}
whose actual forms are given as follows:
\begin{equation}\label{eq72}
\Lambda_0 = \frac{\partial \rho}{\partial t}+\frac{\partial m_r}{\partial x_r}=0
\end{equation}
\begin{equation}\label{eq73}
\Lambda_j=\frac{\partial m_j}{\partial t}+\frac{\partial}{\partial x_r}\left(\frac{m_jm_r}{\rho}+p \delta_{jr}+p_{jr}\right)=0
\end{equation}
\begin{equation}\label{eq74}
p_{jr} \equiv (\ol{\ul{p}}_{jr})_{\rm NS}+\rho \ol{u_j'u_r'}
\end{equation}
\begin{equation}\label{eq75}
\Lambda_4=\frac{\partial E}{\partial t}+\frac{\partial}{\partial x_r}\left(\frac{m_r}{\rho}H+Q_r\right)+\sum_{\rm a,\,b,\,\cdots}e_{\rm a}^0\,W_{\rm a}=0
\end{equation}
\begin{equation}\label{eq76}
Q_r \equiv (\ol{\ul{Q}}_r)_{\rm Fourier}+\rho\ol{u_r'h'}
\end{equation}
\begin{equation}\label{eq77}
\Lambda_{\rm a}=\frac{\partial \rho_{\rm a}}{\partial t}+\frac{\partial}{\partial x_r}\left( \frac{m_r}{\rho}\rho_{\rm a}+M_{{\rm a},\,r}\right)-W_{\rm a}=0
\end{equation}
\begin{equation}\label{eq78}
M_{{\rm a},\,r} \equiv (\ol{\ul{M}}_{{\rm a},\,r})_{\rm Fick}+\rho\ol{u_r'Y_{\rm a}'}
\end{equation}
\begin{equation}\label{eq79}
W_{\rm a}=\frac12 (W_{0 {\rm a}}^+ + W_{0 {\rm a}}^- )
\end{equation}
where $p_{jr}$, $Q_r$, $M_{{\rm a},\,r}$ and $W_{\rm a}$ are average
transports of
momentum, thermal energy, species mass due to diffusion and chemical
reactions, viewed from the center of gravity frame of reference, respectively.
Of the four transport processes three but chemical reactions have a structure
such that molecular and turbulent transports are additive, whereas these
are tightly coupled and built in a single exponential form for chemical
reactions (Table 1). A few remarks are in order in deriving Eqs.(72) through
(78) from (71):
They are exact to $O(Z^{'2})$ if Mach number is small enough. Single term
turbulence corrections as derived here would not be available unless one
followed rule I plus the following one (rule II) that fluctuation of each
density-proportional quantity be written in density-independent quantities
except fluctuation in density itself, for example, $\ul{\ol{m}}_j=m_j+\rho u_j'+\rho' u_j+\rho' u_j'-\ol{\rho' u_j'}$.
It is remarkable that turbulence correction
to each of Navier-Stokes, Fourier and Fick laws is represented by such a
simple expression\cite{b19}
without employing the so-called mass average concept. Note also that artifice
of this kind is not necessary to reach the same result
if phenomenologies are discarded and the
whole deduction to fluid equations starts from the B-formalism in the
phase space, which, unfortunately, is practicable only to nonreactive
monatomic gases.

Equations governing turbulent transports to close the system
emerge from the following
equations\cite{b15}
\begin{equation}\label{eq80}
\ol{Z_\alpha' \what{\Lambda}_\beta' + \Lambda_\alpha' \what{Z}_\beta'} =0,~~((\alpha,\,\beta)\,;\,(0,\,j,\,4,\,{\rm a},\,{\rm b},\,\cdots))
\end{equation}
where $\Lambda_\alpha' = \ul{\Lambda}_\alpha - \Lambda_\alpha$, and
$Z_\alpha'$ denotes fluctuation of fluid variable $Z_\alpha$.
Regardless of the fact that those equations are nonlinear, they are again
separable exactly in the form of (22) and (23), resulting in the following
set of equations (see Appendix for its derivation);
\begin{equation}\label{eq81}
Dq_0+\partial_rq_r=0
\end{equation}
\begin{equation}\label{eq82}
\hspace*{-3mm} \left. \begin{array}{l}
Dq_j+\partial_r(u_r^\dagger q_j +q_{40}\delta_{rj}+q_{rj})+q_r\partial u_j^\dagger /\partial x_r-\rho^{-1}(\partial p/\partial x_j)q_0=0 \vspace*{2mm}\\
q_{jr}=-\mu [\partial_j (\rho^{-1} q_r)+\partial_r(\rho^{-1}q_j)-(2/3)\delta_{jr}\partial_k(\rho^{-1}q_k)] \vspace*{1mm}\\
\hspace*{11mm} \displaystyle -[\partial u_r^\dagger /\partial x_j+\partial u_j^\dagger /\partial x_r-(2/3)\delta_{jr} \partial u_k^\dagger /\partial x_k]\frac{q_4}{\rho R} \frac{d\mu}{dT} \vspace*{1mm}\\
\hspace*{11mm}+\rho^{-1}[q_jq_r-(\delta_{jr}/3)q_k^2]
\end{array} \right\}
\end{equation}\vspace*{-2.5mm}
\begin{equation}\label{eq83}
\hspace*{-3mm} \left. \begin{array}{l}
\displaystyle \frac{1}{\gamma-1} Dq_{40}+\partial_r\left(\frac{\gamma}{\gamma-1}q_{40}u_r^\dagger +q_{rkk}\right)+\sum_{\rm a}e_{\rm a}^0\,q_{{\rm a}W}=0 \vspace*{2mm}\\
\displaystyle q_{rkk}=-(\lambda /R)[\partial_r(\rho^{-1}q_4)+p^{-1}q_4 \frac{d\lambda}{dT} \partial RT/ \partial x_r] \vspace*{1mm}\\
\hspace*{12mm}+[\gamma/(\gamma-1)](RTq_r+\rho^{-1}q_rq_4)
\end{array}\right\}
\end{equation}\vspace*{-2.5mm}
\begin{equation}\label{eq84}
\hspace*{-3mm} \left. \begin{array}{l}
Dq_{\rm a}+\partial_r(u_r^\dagger q_{\rm a}+q_{{\rm a},r})+q_r\partial Y_{\rm a}/\partial x_r-q_{W{\rm a}}+\rho^{-1}q_0W_{\rm a}=0 \vspace*{2mm}\\
\displaystyle q_{{\rm a},r}=-\rho D_{\rm a}\partial_r (\rho^{-1}q_{\rm a})-\frac{d(\rho D_{\rm a})}{dT}\frac{q_4}{\rho R}\frac{\partial Y_{\rm a}}{\partial x_r}+\rho^{-1}q_{\rm a}q_r \vspace*{2mm}\\
q_{W{\rm a}}=\pm m_{\rm a}q_W ~~~~~~~~~~(\pm\,;\,{\rm species~[a]~produced/disappeared})\vspace*{2mm}\\
\displaystyle q_W = \frac{S}{2}\left[ W_0^+-W_0^-+\frac{S}{2}(W_0^++W_0^--W_0^{+-})\right]
\end{array} \right\}
\end{equation}
\begin{equation}\label{eq85}
q_{40}=q_4+\rho R q_0
\end{equation}
where
\begin{eqnarray}
\partial_j \equiv \partial /\partial x_j+\partial /\partial s_j \nonumber
\end{eqnarray}
\begin{eqnarray}
Dq_\alpha \equiv \partial q_\alpha /\partial t-V_{pr}\partial q_\alpha /\partial s_r \nonumber
\end{eqnarray}
Of these equations terms of fluctuations in transport processes
are listed in Table 2.

Turbulent transports having appeared in equations for the average are written
in terms of $q$'s as
\begin{equation}\label{eq86}
\left. \begin{array}{l}
\displaystyle \rho \ol{u_j'u_r'} = \frac{1}{\rho(2\pi)^3}\int_{-\infty}^\infty q_jq_r\,d\vs \vspace*{2mm}\\
\displaystyle \rho \ol{h'u_r'}=\frac{1}{\rho(2\pi)^3} \frac{\gamma}{\gamma -1}\int_{-\infty}^\infty q_4q_r\, d\vs \vspace*{2mm}\\
\displaystyle  \rho \ol{Y_{\rm a}'u_r'} = \frac{1}{\rho(2\pi)^3}\int_{-\infty}^\infty q_{\rm a} q_r\, d\vs \vspace*{2mm}\\
\displaystyle (\ol{T^{'2}})^{1/2} = \frac{1}{\rho R}\left(\int_{-\infty}^\infty q_4^2 \,d\vs \right)^{1/2}
\end{array} \right\}
\end{equation}
Thus the two sets of equations for the average (Eqs.(71)) and separated equations for
fluctuation-correlations (Eqs.(80)) are coupled through those quantities. They have
to be solved simultaneously subject to homogeneous boundary conditions for
$q$'s as
\begin{equation}\label{eq87}
\left. \begin{array}{l}
q_\alpha (\vx_{\rm b},\,\vs)=0~~~~(\alpha \ne 40) \vspace*{2mm}\\
(\partial q_{40}/\partial x_n)_{\vx=\vx_{\rm b}}=0
\end{array} \right\}
\end{equation}
at solid boundaries and at laminar-turbulent flow
boundaries $\vx=\vx_{\rm b}$, also
\begin{equation}\label{eq88}
q_\alpha(\vx,\,\pm\infty)=0
\end{equation}
a necessary condition to secure convergence of the integral in the $\vs$-space.

\section{Concluding remarks}
So far the `Reynolds average' concept is considered not to be applicable to
turbulent combustion gasdynamics. This negative view has its origin in poor
predictivity for one of the four average turbulent transports, namely, the
chemical reaction rate. It is concluded that if one employs bimodal Arrhenius
law to replace the classical one, sound `average' description of turbulent
combustion may well be expected despite its on-off structure of signals
to be observed at the reaction front, a sign of marginal variance amplitude
allowed for statistical description.

The bimodal law as the average turbulent reaction rate and its variance are
derived on the basis of the turbulent equilibrium, an exact solution of
non-equilibrium statistical mechanics. It is only through the microscopic
approach that the correct estimate for population of high energy (reactive)
molecules is possible, and thereby high reactivity under turbulent
environment is elucidated.

Also proposed are Reynolds averaged gasdynamic equations for turbulent
combustion with those renovations incorporated. They are coupled with another
set governing turbulent fluctuations through the turbulent transport
processes to constitute a closed system.

\newpage
\begin{appendix}
\section*{Appendix.~~Equations governing turbulent fluctuation-correlations}
\setcounter{section}{1}
The actual expression for $\Lambda_\alpha'$ of Eq.(80) subject to rules I and II are
\begin{equation}\label{a01}
\left. \begin{array}{l}
\displaystyle \Lambda_0' \equiv \partial \rho'/\partial t+(\partial /\partial x_r)(\rho u_r'+\rho'u_r^\dagger+\rho'u_r'-\ol{\rho'u_r'})=0 \vspace*{2mm}\\
\displaystyle u_r^\dagger = m_r/ \rho =(\rho u_r+\ol{\rho' u_r'})/ \rho
\end{array} \right\}
\end{equation}

\begin{equation}\label{a02}
\left. \begin{array}{l}
\displaystyle \Lambda_j' \equiv \frac{\partial}{\partial t}(\rho' u_j^\dagger+\rho u_j')+\frac{\partial}{\partial x_r}[\rho u_r^\dagger u_j'+\rho u_j^\dagger u_r' \vspace*{1mm}\\
\hspace*{10mm} +\rho'u_j^\dagger u_r^\dagger +\rho u_j'u_r'-\rho\ol{u_j'u_r'}+p'\delta_{jr}+(p'_{jr})_{\rm NS}]=0 \vspace*{4mm}\\
(p_{jr}')_{\rm NS} = -\mu [\partial u_r'/\partial x_j+\partial u_j'/\partial x_r-(2 \delta_{jr}/3)\partial u_k'/\partial x_k] \vspace*{1mm}\\
\hspace*{18mm} -(d\mu /dT)T'[\partial u_r^\dagger /\partial x_j+\partial u_j^\dagger /\partial x_r-(2\delta_{jr}/3)\partial u_k^\dagger /\partial x_k]
\end{array} \right\}
\end{equation}
\begin{equation}\label{a03}
\left. \begin{array}{l}
\displaystyle \Lambda_4' = \frac{1}{\gamma -1}\frac{\partial p'}{\partial t}+\frac{\partial }{\partial x_r} \left[ \frac{\gamma}{\gamma -1}(p' u_r^\dagger +pu_r'+p'u_r'-\ol{p'u_r'}) \right. \vspace*{2mm}\\
\hspace*{10mm} \displaystyle \left. \frac{}{}+(Q_r')_{\rm Fourier}\right] +\sum_{\rm a} e_{\rm a}^0 W_{\rm a}'=0 \vspace*{3mm}\\
(Q_r')_{\rm Fourier}=-\lambda \partial T'/\partial x_r -(d\lambda/dT)T'\partial T/\partial x_r
\end{array} \right\}
\end{equation}
\begin{equation}\label{a04}
\left. \begin{array}{l}
\displaystyle \Lambda_{\rm a}' = \frac{\partial  \rho Y_{\rm a}'}{\partial t} + \frac{\partial}{\partial x_r}\left[ \rho u_r^\dagger Y_{\rm a}' + \rho Y_{\rm a}' u_r' -\rho \ol{Y_{\rm a}'u_r'}+(M_{{\rm a} r}')_{\rm Fick} \right] \vspace*{1mm}\\
\hspace*{10mm} \displaystyle +\rho u_r' \frac{\partial Y_{\rm a}}{\partial x_r}-W_{\rm a}' +\frac{\rho'}{\rho} W_{\rm a}=0 \vspace*{4mm}\\
\displaystyle (M_{{\rm a}r}')_{\rm Fick} = -\rho D_{\rm a} \frac{\partial Y_{\rm a}'}{\partial x_r}-\frac{d(\rho D_{\rm a})}{dT} T'\frac{\partial Y_{\rm a}}{\partial x_r}
\end{array} \right\}
\end{equation}

Eqs.(80) are again separable into respective independent variables $\vx$ and
$\what{\vx}$ with full nonlinear terms retained when subjected to the
following separation rules:
\begin{equation}\label{a05}
\left. \begin{array}{l}
\displaystyle \ol{Z_\alpha' \what{Z}_\beta'} = {\rm R.P.}l^3 \int_{-\infty}^\infty d\vk\, g_\alpha(\vk)\, \what{g}_\beta^*(\vk) \vspace*{2mm}\\
\displaystyle \ol{Z_\alpha' \what{Z}_\beta' Z_\gamma'}={\rm R.P.}l^6\int_{-\infty}^\infty d\vk\, \what{g}_\beta^*(\vk)\int_{-\infty}^\infty d\what{\vk}\, g_\alpha(\vk-\what{\vk})\,g_\gamma (\what{\vk}) \vspace*{2mm}\\
\displaystyle \ol{Z_\alpha' \what{Z}_\beta' \what{Z}_\gamma'} ={\rm R.P.}l^6\int_{-\infty}^\infty d\vk\, g_\alpha (\vk) \int_{-\infty}^\infty d\what{\vk}\, \what{g}_\beta^* (\vk-\what{\vk})\, \what{g}_\gamma^*(\what{\vk})
\end{array} \right\}
\end{equation}
This separation rule preserves the property that periodic part in $g$'s be
separated out after (52), namely,
$$
g_\alpha (\vx,\, \vk) = e^{i\vk\cdot\vx} G_\alpha (\vx,\,\vk)
$$
which converts Eqs.(78) in 6D space $(\vx,\,\what{\vx})$ into those in
another 6D space $(\vx,\vk)$ as
\begin{equation}\label{a06}
D(\vk)G_0+\partial_r(\vk)G_r=0
\end{equation}
\begin{equation}\label{a07}
\hspace*{-3mm} \left. \begin{array}{l}
D(\vk)G_j+\partial_r(\vk)(u_r^\dagger G_j+ G_{40}\delta_{rj}+G_{rj}) \vspace*{1mm}\\
\hspace*{20mm}+(\partial u_j^\dagger /\partial x_r)G_r-\rho^{-1}(\partial p/\partial x_j)G_0=0 \vspace*{2mm}\\
G_{rj}=-\mu [ \partial_j(\vk)(\rho^{-1}G_r)+\partial_r(\vk)(\rho^{-1}G_j)-(2/3)\delta_{jr}\partial_k(\vk)(\rho^{-1}G_k)]  \vspace*{1mm}\\
\hspace*{11.5mm}  -\displaystyle [\partial u_r^\dagger /\partial x_j+\partial u_j^\dagger /\partial x_r  -(2/3)\delta_{jr}\partial u_k^\dagger /\partial x_k] \frac{G_4}{\rho R}\frac{d\mu}{dT}  \vspace*{1mm}\\
\hspace*{11.5mm} +(1/\rho)[\Gamma (G_jG_r)-(\delta_{jr}/3)\Gamma(G_kG_k)]
\end{array} \right\}
\end{equation}
\begin{equation}\label{a08}
\hspace*{-3mm} \left. \begin{array}{l}
\displaystyle \frac{1}{\gamma-1}D(\vk)G_{40}+\partial_r(\vk) \left( \frac{\gamma}{\gamma -1}G_{40} u_r^\dagger +G_{rkk}\right)+\sum_{\rm a}e_{\rm a}^0\,G_{{\rm a}W}=0 \vspace*{2mm}\\
\displaystyle G_{rkk}=-(\lambda/R)[\partial_r(\vk)(\rho^{-1}G_4)+p^{-1}G_4 \frac{d\lambda}{dT} \partial RT/\partial x_r] \vspace*{1mm}\\
\hspace*{14mm}+[\gamma /(\gamma-1)][RTG_r+\rho^{-1}\Gamma(G_rG_4)]
\end{array} \right\}
\end{equation}
\begin{equation}\label{a09}
\hspace*{-3mm} \left. \begin{array}{l}
D(\vk)G_{\rm a}+\partial_r(\vk)(u_r^\dagger G_{\rm a}+G_{{\rm a},r})
 \displaystyle +G_r\partial Y_{\rm a}/\partial x_r-G_{W{\rm a}}+\frac{G_0}{\rho}W_{\rm a}=0 \vspace*{2mm}\\
\displaystyle G_{{\rm a},r}=-\rho D_{\rm a}\partial_r(\vk)(\rho^{-1}G_{\rm a})-\frac{d(\rho D_{\rm a})}{dT} \frac{G_4}{\rho R}\frac{\partial Y_{\rm a}}{\partial x_r}+\rho^{-1}\Gamma (G_{\rm a}G_r)
\end{array} \right\}
\end{equation}
where we have defined the following quantities,
\begin{equation}\label{a10}
D(\vk)G \equiv \partial G/\partial t-i\vk\!\cdot\!\mvec{V}_pG
\end{equation}
\begin{equation}\label{a11}
\partial_r(\vk) \equiv \partial /\partial x_r+ik_r
\end{equation}
\begin{equation}\label{a12}
\Gamma(G_\alpha G_\beta) \equiv \int_{-\infty}^\infty G_\alpha(\vk-\vk')G_\beta(\vk')d\vk'
\end{equation}
\begin{equation}\label{a13}
G_{40}=G_4+RTG_0
\end{equation}

Eqs.(A4) through (A7) govern the `wave' functions $G_0,\,G_j,\,G_4(G_{40})$
and $G_{\rm a},\,G_{\rm b},\,\cdots$ from which turbulent correlations
$\ol{Z_\alpha' \what{Z}_\beta'}$ are calculated. Relationships between the
two groups of variables are the following;
\begin{equation}\label{a14}
Z_\alpha'=\left( \begin{array}{c}
\rho' \\ \rho u_j' \\ \rho RT' \\ p' \\ \rho Y_{\rm a}' \\ W_{\rm a}'
\end{array} \right) ~~~~ G_\alpha=\left( \begin{array}{c}
G_0 \\ G_j \\ G_4 \\ G_{40} \\ G_{\rm a} \\ G_{W{\rm a}}
\end{array} \right)
\end{equation}
Turbulent transports that appear in the equations for the average are
\begin{equation}\label{a15}
\hspace*{-7mm}\left. \begin{array}{l}
\displaystyle \rho \ol{u_j'u_r'}=\frac{1}{\rho} {\rm R.P.}l^3 \int_{-\infty}^\infty G_jG_r^* d\vk,~~{\rm (Reynolds~stress)} \vspace*{2mm}\\
\displaystyle \rho \ol{h'u_r'}=\frac{1}{\rho}\frac{\gamma}{\gamma-1}{\rm R.P.}l^3 \int_{-\infty}^\infty G_4G_r^* d\vk,~~{\rm (turbulent~heat~flux~density)} \vspace*{2mm}\\
\displaystyle \rho \ol{Y_{\rm a}'u_r'}=\frac{1}{\rho}{\rm R.P.}l^3 \int_{-\infty}^\infty G_{\rm a}G_r^* d\vk,~~{\rm (turbulent~diffusion)} \vspace*{2mm}\\
\displaystyle (\ol{T^{'2}})^{1/2}=\frac{1}{\rho R}\left( l^3\int_{-\infty}^\infty G_4G_4^* d\vk \right)^{1/2},~~{\rm (turbulent~chemical~reaction)}
\end{array} \right\}
\end{equation}

These equations as derived above in the $\vk$-space are integro-differential
equations having nonlinear integrals $\Gamma$. These integral expressions
are of convolution type, so are eliminated by Fourier transform
\begin{equation}\label{a16}
G_\alpha (\vx,\,\vk)=\frac{1}{(2\pi l)^3}\int_{-\infty}^\infty d\vs\, e^{-i\vk\cdot\vs}q_\alpha(\vx,\,\vs)
\end{equation}
leading to an equivalent set of equations in eddy ($\vs$) space simply
through a transformation rule posted in Table A. Their actual forms are:
\begin{equation}\label{a17}
Dq_0+\partial_rq_r=0
\end{equation}
\begin{equation}\label{a18}
\hspace*{-3mm} \left. \begin{array}{l}
Dq_j+\partial_r(u_r^\dagger q_j+q_{40}\delta_{rj}+q_{rj})+q_r\partial u_j^\dagger /\partial x_r-\rho^{-1}(\partial p/\partial x_j)q_0=0 \vspace*{2mm}\\
q_{jr}=-\mu [\partial_j (\rho^{-1} q_r)+\partial_r(\rho^{-1}q_j)-(2/3)\delta_{jr}\partial_k(\rho^{-1}q_k)] \vspace*{1mm}\\
\hspace*{11mm} \displaystyle -[\partial u_r^\dagger /\partial x_j+\partial u_j^\dagger /\partial x_r-(2/3)\delta_{jr} \partial u_k^\dagger /\partial x_k]\frac{q_4}{\rho R}\frac{d\mu}{dT} \vspace*{1mm}\\
\hspace*{11mm}+\rho^{-1}[q_jq_r-(\delta_{jr}/3)q_k^2]
\end{array} \right\}
\end{equation}
\begin{equation}\label{a19}
\hspace*{-3mm} \left. \begin{array}{l}
\displaystyle \frac{1}{\gamma-1} Dq_{40}+\partial_r\left(\frac{\gamma}{\gamma-1}q_{40}u_r^\dagger +q_{rkk}\right)+\sum_{\rm a}e_{\rm a}^0\,q_{{\rm a}W}=0 \vspace*{2mm}\\
\displaystyle q_{rkk}=-(\lambda /R)[\partial_r(\rho^{-1}q_4)+p^{-1}q_4 \frac{d\lambda}{dT}\partial RT/ \partial x_r] \vspace*{1mm}\\
\hspace*{12mm}+[\gamma/(\gamma-1)](RTq_r+\rho^{-1}q_rq_4)
\end{array}\right\}
\end{equation}
\begin{equation}\label{a20}
\hspace*{-3mm} \left. \begin{array}{l}
Dq_{\rm a}+\partial_r(u_r^\dagger q_{\rm a}+q_{{\rm a},r})+q_r\partial Y_{\rm a}/\partial x_r-q_{W{\rm a}}+\rho^{-1}q_0W_{\rm a}=0 \vspace*{2mm}\\
\displaystyle q_{{\rm a},r}=-\rho D_{\rm a}\partial_r (\rho^{-1}q_{\rm a})-\frac{d(\rho D_{\rm a})}{dT}\frac{q_4}{\rho R}\frac{\partial Y_{\rm a}}{\partial x_r}+\rho^{-1}q_{\rm a}q_r \vspace*{2mm}\\
q_{W{\rm a}}=\pm m_{\rm a}q_W ~~~~~~~~~~(\pm\,;\,{\rm species~[a]~produced/disappeared})\vspace*{2mm}\\
\displaystyle q_W = \frac{S}{2}\left[ W_0^+-W_0^-+\frac{S}{2}(W_0^++W_0^--W_0^{+-})\right]
\end{array} \right\}
\end{equation}
\begin{equation}\label{a21}
q_{40}=q_4+\rho R q_0
\end{equation}
\end{appendix}

\newpage
\section*{Figure captions}
\noindent
Fig.1~~Dependence of flame velocity $u_T/u_L$ on turbulent intensity
$u'/u_L$ of methane-air premixture in elevated pressure environments.
Kobayashi et al (1996)\cite{b03}. \vspace*{4mm}\\
Fig.2~~Failure of the classical theory ($\delta T=0$) in predicting ignition
 time $\tau$ at low-temperatures ($T<1000^\circ K$), as corrected by turbulent
reaction rate formula ($\delta T>0$)\cite{b11}. \vspace*{4mm}\\
Fig.3~~Turbulent flame velocity $u_T/u_L$ as dependent on turbulent intensity
$u'/u_L$ of 9.0 percent $H_2$-air premixture due to a predecessor
theory\cite{b14} of the current one as compared with existing experiments,
also with Arrhenius law equivalent and renormalization
group theory. \vspace*{4mm}\\
Fig.4~~Maxwellian equilibrium $f_0~(\delta T=0)$ and Turbulent equilibrium
$f^0~(\delta T>0)$ distribution function plotted against absolute
molecular velocity, revealing appreciable difference in the population of
high-energy molecules eligible for chemical reactions. \vspace*{4mm}\\
Fig.5~~Relative variance of reaction rates under turbulence around (a)
averaged Arrhenius reaction rate and (b) turbulent (bimodal) reaction rate.

\section*{Table captions}
\noindent
Table 1~~Molecular and turbulent transport processes \vspace*{4mm}\\
Table 2~~Fluctuations in transport processes \vspace*{4mm}\\
Table A~~Conversion rules from wave-number space $(\vk)$ to eddy space $(\vs)$

\newpage
\begin{center}
\def\arraystretch{2.5}
\begin{tabular}{|p{40mm}|l|} \hline
\begin{tabular}{l}momentum (shear) \vspace*{-6mm}\\  flux density \end{tabular} & $\displaystyle p_{jl}=-\mu \left(\frac{\partial u_j^\dagger}{\partial x_l}+\frac{\partial u_l^\dagger}{\partial x_j}-\frac23 \delta_{jl}\frac{\partial u_k^\dagger}{\partial x_k}\right)+\rho \ol{u_j'u_l'}$ \vspace*{-10mm}\\
 & ~~~~~~~~(Navier-Stokes' law) ~~~~~~~~~~ (Reynolds stress) \vspace*{-8mm}\\
 & $(u_j^\dagger =m_j / \rho)$ \\ \hline
\begin{tabular}{l}heatflux density \end{tabular} & $\displaystyle q_j=-\lambda \frac{\partial T^\dagger}{\partial x_j}~~~~~~~~~~~~~~+\rho \ol{u_j'h'}$ \vspace*{-6mm}\\
 & ~~~~~~(Fourier's law)~~~~~~~~(turbulent heatflux density) \vspace*{-8mm}\\
 & $(T^\dagger=p/\rho R)$ \\ \hline
\begin{tabular}{l}partial mass flux \vspace*{-6mm}\\ density \end{tabular} & $\displaystyle ~~~\,=-\rho D\frac{\partial Y_{\rm a}^\dagger}{\partial x_j}~~~~~~~~+\rho\ol{u_j'Y_{\rm a}'}$ \vspace*{-10mm}\\
 & ~~~~~~(Fick's law)~~~~~~~~(turbulent diffusion) \vspace*{-8mm}\\
 & $(Y_{\rm a}^\dagger =\rho_{\rm a}/\rho)$ \\ \hline
\begin{tabular}{l}chemical reaction \vspace*{-6mm}\\ rate \end{tabular} & $\begin{array}{l} \displaystyle W=\frac12 \left[ A(n_{\rm a}+\delta n_{\rm a})(n_{\rm b}+\delta n_{\rm b})\exp \left(-\frac{E}{{\cal R}(T+\delta T)} \right) \right.\\
\displaystyle ~~~~~~~~~~\left. +A(n_{\rm a}-\delta n_{\rm a})(n_{\rm b}-\delta n_{\rm b})\exp \left(-\frac{E}{{\cal R}(T-\delta T)} \right) \right] \end{array}$ \vspace*{-4mm}\\
 & ~~~~~~(turbulence-corrected Arrhenius law) \\ \hline
\end{tabular}
\vspace{4mm} \\Table 1
\end{center}

\newpage
\begin{center}
\def\arraystretch{2.5}
\hspace*{-2mm}
\begin{tabular}{|l||l|l|} \hline
~fluctuations in & ~molecular & ~turbulent \\ \hline
\begin{tabular}{l}momentum (shear) \vspace*{-6mm}\\  flux density \end{tabular} & $\begin{array}{l}-\mu [ \partial_j (\rho^{-1} q_r)+\partial_r(\rho^{-1}q_j) \vspace*{-6mm}\\ ~-(2/3)\delta_{jr}\partial_k(\rho^{-1}q_k)] \vspace*{-6mm}\\  ~-[\partial u_r^\dagger /\partial x_j+\partial u_j^\dagger /\partial x_r \vspace*{-6mm}\\ \displaystyle ~-(2/3)\delta_{jr} \partial u_k^\dagger /\partial x_k]\frac{q_4}{\rho R}\frac{d\mu}{dT} \end{array}$ & $+\rho^{-1}[q_jq_r-(\delta_{jr}/3)q_k^2]$ \\ \hline
\begin{tabular}{l}heatflux density \end{tabular} & $\begin{array}{l} -(\lambda /R)[\partial_r(\rho^{-1}q_4) \vspace*{-6mm}\\ \displaystyle ~+p^{-1}q_4 \frac{d\lambda}{dT} \partial RT/ \partial x_r] \end{array}$ & $+[\gamma/(\gamma-1)](RTq_j+\rho^{-1}q_jq_4)$ \\ \hline
\begin{tabular}{l}partial mass flux \vspace*{-6mm}\\ density \end{tabular} & $\begin{array}{l} -\rho D_{\rm a}\partial_r (\rho^{-1}q_{\rm a}) \vspace*{-4mm}\\ \displaystyle ~-\frac{d(\rho D_{\rm a})}{dT}\frac{q_4}{\rho R}\frac{\partial Y_{\rm a}}{\partial x_r} \end{array} $ & $+\rho^{-1}q_{\rm a}q_r$ \\ \hline
\begin{tabular}{l}chemical reaction \vspace*{-6mm}\\ rate \end{tabular} & \multicolumn{2}{c|}{$\displaystyle \frac{S}{2}\left[ W_0^+-W_0^- +\frac{S}{2}(W_0^++W_0^--W_0^{+-})\right]$} \\ \hline
\end{tabular}
\vspace{4mm} \\Table 2
\end{center}

\newpage
\begin{center}
\def\arraystretch{2.5}
\begin{tabular}{|lcl|} \hline
~~wave number ($\vk$) space & $\rightarrow$ & eddy $(\vs)$ space \\ \hline
~~$G_\alpha(\vx,\,\vk)$ & $\rightarrow$ & $q_\alpha(\vx,\,\vs)$ \\
~~$ik_j$ & $\rightarrow$ & $\partial / \partial s_j$ \\
~~$\partial_j(\vk)=\partial/\partial x_j+ik_j$ & $\rightarrow$ & $\partial_j=\partial /\partial x_j+\partial /\partial s_j$ \\
~~$D(\vk)G_\alpha=\partial G_\alpha/\partial t-i\vk\!\cdot\!\mvec{V}_pG_\alpha$ & $\rightarrow$ & $Dq_\alpha=\partial q_\alpha/\partial t-V_{pr}\partial q_\alpha /\partial s_r$ \\
~~$\displaystyle l^3\!\int_{-\infty}^\infty G_\alpha(\vk')G_\beta(\vk-\vk')d\vk'$ & $\rightarrow$ & $q_\alpha q_\beta$ \\
\hline
\end{tabular}
\vspace*{4mm}\\Table A
\end{center}

\end{document}